\documentclass[smallextended]{svjour3}
\usepackage{graphicx}
\usepackage{rotating}
\usepackage{amssymb}
\usepackage{mathptmx}
\usepackage{enumitem}
\usepackage{amsmath}
\usepackage{bm}
\makeatletter
\journalname{Journal of Low Temperature Physics}
\usepackage[
style=numeric,
sorting=none,
giveninits=true,
maxbibnames=8,
date=year,
isbn=false,
doi=false,
url=false,
]{biblatex}
\DeclareNameAlias{author}{family-given}
\renewbibmacro{in:}{}
\AtEveryBibitem{\clearfield{number}}
\AtBeginBibliography{\small}
\DeclareFieldFormat[article]{volume}{\mkbibbold{#1}}

\DeclareFieldFormat[article]
{pages}{#1}
\DeclareFieldFormat[inproceedings, incollection, inbook]
{pages}{p. #1}

\newcommand{\mat}[1]{\ensuremath{\mathbf #1}}   % Bold upright for tensors
\newcommand{\matT}[1]{\ensuremath{\mathbf #1}^\mathrm{T}}   % Bold upright for tensors
\newcommand{\inv}[1]{\ensuremath{{\mathbf #1}^{-1}}}   % Bold upright for tensors
\renewcommand{\vec}[1]{\mat{#1}}
\newcommand{\vecT}[1]{\matT{#1}}

\newcommand{\dif}    {\ensuremath{\,\mathrm{d}}} % Math differential qty

\newcommand{\be}{\begin{equation}}
\newcommand{\ee}{\end{equation}}
\newcommand{\ba}{\begin{align}}
\newcommand{\ea}{\end{align}}

\begin{document}

\title{Energy calibration of nonlinear microcalorimeters with uncertainty estimates from Gaussian process regression}
\titlerunning{Energy calibration uncertainty from Gaussian process regression}

\author{J.W.~Fowler   \and B.~K.~Alpert \and G.~C.~O'Neil \and D.~S.~Swetz \and J.~N.~Ullom}

\institute{All authors \at National Institute of Standards and Technology, 325 Broadway, Boulder, Colorado 80305, USA. 
\and
Fowler \and Ullom \at University of Colorado, Department of Physics, Boulder, Colorado 80309, USA. \\
%Tel: +1 303-497-3990.\\
\email{joe.fowler@nist.gov}
}

\vspace{-2cm}
 \date{Draft of: \today}
\maketitle
\vspace{-5mm}

%%%%%%%%%%%%%%%%%%%%%%%%%%%%%%%%%%%%%%%%%%%%%%%%
\begin{abstract}

The nonlinear energy response of cryogenic microcalorimeters is usually corrected through an empirical calibration. X-ray or gamma-ray emission lines of known shape and energy anchor a smooth function that generalizes the calibration data and converts detector measurements to energies. We argue that this function should be an approximating spline. The theory of Gaussian process regression makes a case for this functional form. It also provides an important benefit previously absent from our calibration method: a quantitative uncertainty estimate for the calibrated energies, with lower uncertainty near the best-constrained calibration points.


\end{abstract}

%%%%%%%%%%%%%%%%%%%%%%%%%%%%%%%%%%%%%%%%%%%%%%%%
\section{Introduction}
%%%%%%%%%%%%%%%%%%%%%%%%%%%%%%%%%%%%%%%%%%%%%%%%

Cryogenic microcalorimeters have been used to measure x-ray and gamma-ray emission in many diverse settings. Considering only existing x-ray spectrometers based on the transition-edge sensor (TES), these applications include~\cite{Doriese:2017} synchrotron beamlines, astrophysical telescopes, exotic atom measurements at particle accelerators, computed tomography with spectral discrimination, and even metrological study of fluorescence energies. One requirement these applications share is accurate energy estimation for each photon detected, a difficult challenge given that the measurements are made with sensors whose energy response is inherently nonlinear. 

Users of such devices generally employ an empirically estimated \emph{calibration curve}, a function that converts each measured pulse height to a photon energy. The pulse height is typically  estimated by statistically optimal filtering, corrected for confounding effects such as slow drifts in system gain or for a bias that depends on the relative arrival time of a photon and the system's sampling clock~\cite{Fowler2016}. We will call this optimal, corrected value simply the \emph{pulse height} (PH, or $p$ in equations). The calibration curve is ``anchored'' by a small number of spectral features. These features must have an absolute energy that is known \emph{a priori}, and they must be detected with high enough intensity that their representative pulse height can be estimated from the measured spectrum with small uncertainties. Such features might be gamma rays, x-ray fluorescence line emission, or light emitted by a calibrated x-ray monochromator.

To create a calibration curve from a set of anchor points, one must make several decisions. Should the curve interpolate or approximate the data? Should it be a polynomial, a spline, or another functional form? How should parameters such as the degree of the polynomial or the number and locations of the spline's knots be chosen? Should the smoothing function relate energy to PH directly or indirectly?

We think the answer to the first question is clear: it is best to have a calibration curve approximate the data. Anchor points have uncertainty in both the PH and energy, and only an approximating curve can fully account for the uncertainties. With polynomial fits, approximation is accomplished by weighted least-squares fits to a polynomial of degree lower than the number of anchor points. Though splines normally interpolate their data, several methods to approximate with splines exist; we have previously found best results with a cubic smoothing spline~\cite{Fowler2021}.

Our most important result is the derivation of the \emph{uncertainty} on the calibration function from the theory of Gaussian process regression (GPR). It quantifies our intuition that the energy scale is more uncertain at energies that are further from the anchor points, or near to anchor points that themselves have higher uncertainty. GPR also justifies the smoothing spline form.

% Generalize beyond TESs?
The procedure described here has been tested only with data from TES microcalorimeters with Mo-Cu bilayers and designed with normal-metal banks and bars. Such sensors have nonlinear energy responses that are difficult or impossible to model a priori, and empirical calibrations are necessary. We expect that the GPR procedure should generalize to other TES designs or other sensors, but without testing other cases, we cannot state when it will be needed.

%%%%%%%%%%%%%%%%%%%%%%%%%%%%%%%%%%%%%%%%%%%%%%%%
\section{Choosing a calibration space} \label{sec:cal_space}
%%%%%%%%%%%%%%%%%%%%%%%%%%%%%%%%%%%%%%%%%%%%%%%%

The calibration curve is a function that estimates the photon energy from a PH value $p$. It can employ a spline directly or indirectly. Indirect use means the spline models some $y(E,p)$ as a function of some $x(p)$, so long as $x(\cdot)$ is a monotone function of its input and $y(E,p)$ is a function for which $y_0=y(E,p)$ can be easily solved for $E$ given $y_0$ and $p$. Many functions $x$ and $y$, or \emph{calibration spaces} are possible. Here we consider $p$ and $\log p$ for $x(p)$; and $E$, $p/E$, $E/p$, and $\log E$ for $y(E,p)$. 

\begin{table}[]
    \centering
    \begin{tabular}{rr|rrrr}
    \multicolumn{2}{c}{Calibration space} & \multicolumn{4}{c}{Median linear-interpolation error (eV)} \\
       $x(p)$ & $y(E,p)$ & Ti/Cr/Fe & V/Mn/Co & Cr/Fe/Ni & Mn/Co/Cu  \\ \hline
        $p$ & $E$ & 16.8 & 18.8 & 20.2 & 22.5\\
        % $p$ & $p/E$ & 0.48 & 0.53 & 0.72 & 1.32 \\
        $p$ & $p/E$ & 0.5 & 0.5 & 0.7 & 1.3 \\
        $p$ & $E/p$ & 2.1 & 2.6 & 3.0 & 4.0 \\
        $p$ & $\log(p/E)$ & 1.3 & 1.6 & 1.9 & 2.7 \\
        % $\log p$ & $E$  &  91.4 & 92.6 & 93.6 & 95.2 \\ 
        $\log p$ & $p/E$ &  7.9 & 8.7 & 9.3 & 10.5 \\
        $\log p$ & $E/p$ &  9.5 & 10.7 & 11.6 & 13.3 \\
        $\log p$ & $\log E$  &  8.7 & 9.7 & 10.5 & 11.9 \\ 
        
    \end{tabular}
    \caption{Median energy error across all TESs when the linear interpolation of two points is used at an intermediate point, for seven calibration spaces. The column ``Ti/Cr/Fe'' uses linear interpolation of the K$\alpha$ lines of Ti and Fe to estimate the Cr K$\alpha$ energy, and similarly for the last three columns. Each interpolation spans approximately 1\,keV\@. The intermediate elements have K$\alpha$ energies of 5.4\,keV (Cr), 5.9\,keV (Mn), 6.4\,keV (Fe), and 6.9\,keV (Co). The simplest calibration space (row 1, $E$ vs.\ $p$) introduces large errors. The smallest errors arise when gain $p/E$ vs.\ $p$ is taken to be linear (row 2). The use of $x(p)=\sqrt{p}$ (not listed) gives results intermediate between $p$ and $\log p$.  In any calibration space, the linear-interpolation error is larger at higher photon energy for these data (from a study of lanthanide metal emission~\cite{Fowler2021}).
    \label{tab:cal_space}
    }
\end{table}

The $x$ and $y$ functions can be picked by the principle that the spline function should ``do the least work.'' For example, in the case of TES calibration from 4\,keV to 10\,keV for a study of rare-earth L lines~\cite{Fowler2021}, we found it best to spline $y(E,p)=p/E$, which we called the sensor \emph{gain}, as a function of $x(p)=p$. Consider the K$\alpha$ lines of Mn, Co, and Cu at approximately 6\,keV, 7\,keV, and 8\,keV\@. We asked how far in error the intermediate (Co K$\alpha$) energy estimate would be if the curve were the linear interpolation of the Mn and Cu K$\alpha$ peaks for each possible choice of $y$ and $x$ functions. Only $p/E$ vs.\ $p$ was correct to within 1.3\,eV for the median detector; linear interpolations of other calibration spaces yielded errors up to 23\,eV (Table~\ref{tab:cal_space}). It is possible that other calibration spaces would be preferred for other measurements, because the least-curvature space potentially depends on any of: the detector material and shape, its bias voltage, cryogenic bath temperature, and the energy range of interest. The 3-point test of linear-interpolation error offers a simple way to choose. A second benefit of choosing $x=p$ and $y=p/E$ (or $y=E/p$ or $y=\log(p/E)$) is that any finite, positive value of $y$ when $x=p=0$ will guarantee that the calibration curve yields the expected energy $E=0$ for pulses of size $p=0$.

%%%%%%%%%%%%%%%%%%%%%%%%%%%%%%%%%%%%%%%%%%%%%%%%
\section{Smoothing splines for approximation} \label{sec:smoothing_splines}
%%%%%%%%%%%%%%%%%%%%%%%%%%%%%%%%%%%%%%%%%%%%%%%%

In previous work~\cite{Fowler2021,Fowler2017}, we argued that the smoothing spline is the best way to generalize the calibration $x$-$y$ relationship.  It does not interpolate the anchor points exactly but strikes a compromise between fidelity to the data and minimal curvature. We assume that the more a spline has to curve, the poorer a model it is for calibration---particularly if the $y$ and $x$ functions are chosen to require minimal curvature, as proposed in the previous section. Of all twice-differentiable functions, a cubic smoothing spline is the one~\cite{Green1994} that minimizes the penalty functional (``cost function'')
\begin{equation}
    C[h] \equiv \sum_{i=1}^n\,\left(\frac{h(x_i)-y_i}{\sigma_i}\right)^2 + \lambda\int_a^b\,|h''(x)|^2 \label{eq:cost}\dif x
\end{equation}
where $x_1\ldots x_n$ and $y_1\ldots y_n$ are the anchor points, and $\sigma_i$ is the uncertainty on $y_i$. The first term in $C$ is the usual $\chi^2$ statistic for disagreement between data and model, as we expect $h(x_i)\approx y_i\pm\sigma_i$. The integral is the model's curvature over the range $a=\mathrm{min}(x_i)$ to $b=\mathrm{max}(x_i)$. A regularization parameter $\lambda$ controls the balance between data fidelity and minimization of curvature. For $\lambda=0$, curvature is not penalized, and minimization of $C$ produces an interpolating spline. In the limit $\lambda\rightarrow\infty$, curvature is forbidden, and $h$ will be a line: the line that minimizes the (uncertainty-weighted) sum of squared error between data and model.\footnote{When the data can be exactly interpolated by a line,  that line is found for any value of $\lambda$.}
For finite $\lambda$, a cubic\footnote{Defining curvature as the integral of the $k$th derivative squared yields~\cite{Wahba1978} splines of degree ($2k-1$).}  smoothing spline results.
It is a cubic spline function $h(x)$ with $n$ knots at $x_i$ and natural boundary conditions---that is, $h''(x)=0$ at the lowest and highest values of $x_i$.

This method raises two important questions. First, what curvature penalty $\lambda$ is appropriate? For Gaussian errors, the first term in Equation~\ref{eq:cost} has expected value $E[\chi^2]=n$. A $\lambda$ is reasonable if it yields $\chi^2\approx n$ when cost $C$ is minimized, but we could use a more principled approach. Second, what is the calibration uncertainty on the minimum-cost curve? Gaussian process regression answers both questions.

%%%%%%%%%%%%%%%%%%%%%%%%%%%%%%%%%%%%%%%%%%%%%%%%
\section{Calibration curves as a Gaussian process} \label{sec:gaussian_process}
%%%%%%%%%%%%%%%%%%%%%%%%%%%%%%%%%%%%%%%%%%%%%%%%

% Brad's draft 9/29, again edited by me
A Gaussian process (GP) model describes a distribution over functions.  It posits that the distribution of function \emph{values} at any finite set of points \vec{x} in the domain is a multivariate Gaussian, characterized by its mean and covariance.  It generalizes the mean from a vector to a function $m(x)$ and the covariance from a matrix to a function of two variables $k(x,x’)$. The initial distribution is consistent with a broad class of models, not informed by the observed data \vec{y}.  After observations,\footnote{Estimates of the uncertainty on the measurements are also required, for which we use the simplest possible model: that the noise is independent and Gaussian-distributed with mean zero and variance $\sigma_i^2$.} the mean and covariance are made to be consistent with the data. The mean now estimates the calibration function $h(x)$ we want to learn; the covariance characterizes its uncertainty.  This formulation is a Bayesian framework, with the a priori distribution modified by the observed data to yield an a posteriori distribution.  The posterior distribution enables computation of expected values and covariances for any set of points \vec{x_\star}.  GPR means the refinement, or regression, of a GP model given the data ~\cite{Murphy2012,Rasmussen2006}.

The GPR analysis of $h$ would immediately solve the second difficulty we have with calibration splines, because GPR provides both a model function---the expected value of $h(x)$---and an uncertainty measure---the expected covariance of $h$ at any two points $(x,x')$. Choosing the mean and covariance functions $m$ and $k$ to specify the model space might at first seem a more open-ended problem than the question that ended the previous section: how much to penalize curvature.

Remarkably, \emph{there is a perfectly plausible GP model space that yields smoothing splines as the expected GPR function values:} those functions that are the integral of a continuous random walk (that is, a once-integrated Wiener process) over the interval bounded by the $x_i$ values plus a line over the entire real domain~\cite{Wahba1978}. Integrating a Wiener process once allows the expected slope to differ at the two ends of the measured interval. Such a GP model has expected value and covariance\footnote{Here the covariance is simplified by assuming the domain is transformed to $[\mathrm{min}\,x_i,\mathrm{max}\,x_i]=[0,1]$.}
\begin{align}
m(x) &= \beta_0+\beta_1 x \hspace{1em}\mathrm{and} \label{eq:m} \\
k(x,x') &= \sigma_f^2[v^3/3+v^2|x-x'|/2] \label{eq:kxxp}
\end{align}
where $v\equiv\min(x,x')$. The two parameters $(\beta_0,\beta_1)$ of the mean are assumed to have a ``diffuse'' or uninformative Bayesian prior, so that all possible lines are equally probable. The parameter $\sigma_f^2$ that scales the covariance function controls the expected amount of curvature; a larger $\sigma_f^2$ corresponds to higher curvature.

This GP space is a reasonable way of modeling functions that are linear on either side of the measured interval (potentially with different slopes), and as little structure as possible inside it. Given that the space is reasonable and yields the same, convenient smoothing spline functions we have already shown to work quite well, it is the model space we use for calibration curves.

What curvature scale $\sigma_f^2$ should be used for the Wiener process (random walk) component in Equation~\ref{eq:kxxp}? This is equivalent to the question, what curvature penalty $\lambda$ should be used in the smoothing spline optimization of Equation~\ref{eq:cost}?
Rasmussen and Williams~\cite{Rasmussen2006} (R\&W) show that our regularization parameter $\lambda$ (Equation~\ref{eq:cost}) and the curvature scale $\sigma_f^2$ (Equation~\ref{eq:kxxp}) are related by $\lambda\sigma_f^2=1$. They also give an expression for the marginal likelihood, i.e.\ the likelihood after integrating (marginalizing) over the values of the spline function at its knots. This marginal likelihood is a function of $\sigma_f^2$ and appears below as Equation~\ref{eq:marglikely} (or as R\&W Equation 2.45). The value of $\sigma_f^2$ that maximizes the marginal likelihood is the value most consistent with the data. 

Equations~\ref{eq:m} and \ref{eq:kxxp} yield a GP model that R\&W call GPR with a basis set.  For measurement points \vec{x} and sample points \vec{x_\star}, we have measurements $\vec{y}$ and sample point values $\vec{h_\star}$ jointly distributed as the multivariate Gaussian
\begin{equation}
\left[\begin{array}{l}\vec{y}\\ \vec{h_\star}\end{array}\right]\sim{\cal N}\left(\beta_0+\beta_1\left[\begin{array}{l}\vec{x}\\ \vec{x_\star}\end{array}\right],\;\left[\begin{array}{ll}\mat{K_\vec{y}}&\mat{K_\star}\\
\matT{K_\star}&\mat{K_{\star\star}}\end{array}\right]\right),\label{eq:joint}
\end{equation}
prior to marginalizing over the observed measurements \vec{y}, where \mat{K} is the GP covariance matrix with $K_{ij}=k(x_i, x_j)$; $\mat{K_y}=\mathrm{diag}(\vec{\sigma}^2)+\mat{K}$ is the measurement noise plus GP covariance; $\vec{\sigma^2}=[\sigma_1^2,\ldots,\sigma_n^2]^\mathrm{T}$; $(K_\star)_{ij} = k(x_i,x_{\star j})$; and $(K_{\star\star})_{ij} = k(x_{\star i},x_{\star j})$. The marginalized, predictive distribution is $\vec{h_\star}|(\vec{x},\vec{y},\vec{x_\star})\sim{\cal N} \left(\overline{\vec{h}}_\star,\;\mathrm{cov}(\vec{h_\star})\right)$ where
\begin{align}
\overline{\vec{h}}_\star\equiv\mathbb{E}[\vec{h_\star}|\vec{x},\vec{y},\vec{x_\star}]&=\vec{m}(\vec{x_\star})+\matT{K_\star}\mat{K_y}^{-1}[\vec{y}-\vec{m}(\vec{x})]\label{eq:margmean}\\
\mathrm{cov}(\vec{h_\star})&=\mat{K_{\star\star}}-\matT{K_\star}\mat{K_y}^{-1}\mat{K_\star}\label{eq:variance}
\end{align}
with $\vec{m}(\vec{x})=\beta_0+\beta_1\vec{x}$.

In Equations~\ref{eq:m} and \ref{eq:kxxp}, the parameters $\beta=\langle\beta_0,\beta_1\rangle^\mathrm{T}$ and $\sigma_f^2$ were given, but we want to choose them to maximize the likelihood of the measurements \vec{y}.  Equation~\ref{eq:joint} shows that the non-zero mean of the GP is linear in the parameters $\beta$. Suppose a regression will be done with $n$ calibration anchor points and two basis functions: the pair $h_0(x)=1$ and $h_1(x)=x$. Let \mat{H} be the $2\times n$ matrix whose rows are the basis functions at the $n$ locations \vec{x}. For $\beta\sim{\cal N}(\vec{b},\mat{B})$, minimization of the expected error with respect to $\beta$ yields $\overline{\beta}=\left(\mat{H}\mat{K_y}^{-1}\matT{H}+\mat{B}^{-1}\right)^{-1}\left(\mat{H}\mat{K_y}^{-1}\vec{y}+\mat{B}^{-1}\vec{b}\right)$.  An uninformative Bayesian prior implies $\mat{B}^{-1}\rightarrow\mat{0}$, for which $\vec{b}$ becomes irrelevant, and
\begin{equation}
\overline{\beta}=\left(\mat{H}\mat{K_y}^{-1}\matT{H}\right)^{-1}\mat{H}\mat{K_y}^{-1}\vec{y}.
\end{equation}
Wahba~\cite{Wahba1978} has shown that the expected function values $\overline{\vec{h}}_\star$ (Equation~\ref{eq:margmean}), with $\beta=\overline{\beta}$, is a cubic spline of $\vec{x_\star}$ with knots at the measurement locations $\vec{x}$ and with natural boundary conditions. Therefore, we can use a shortcut. Instead of evaluating the function for every value of $x_\star$ of interest, we can evaluate it only $n$ times---by choosing sample points $\vec{x_\star}$ to coincide with the knots $\vec{x}$. The approximating function we seek is the unique cubic spline with natural boundary conditions that interpolates these $n$ predictions (which are near to but not exactly the values \vec{y}).

To use the above equations, we need the factor $\sigma_f^2$ that scales the covariance function $k(x,x')$ and thus every entry in $\mat{K}$, $\mat{K}_\star$, and $\mat{K}_{\star\star}$. We set  $\sigma_f^2$ to the value that maximizes the marginal likelihood (the Bayesian probability of measuring \vec{y}), obtained by a somewhat elaborate calculation described in R\&W:
\begin{equation}
2\log P\left(\vec{y}|\vec{x},\sigma_f^2\right) = -\vecT{y}\mat{K_y}^{-1}\vec{y}+\vecT{y}\mat{C}\vec{y}-\log|\mat{K_y}|-\log|\mat{A}|-(n-2)\log 2\pi\label{eq:marglikely}
\end{equation}
where $\mat{A}\equiv\mat{H}\inv{K_y}\matT{H}$ and  $\mat{C}\equiv\inv{K_y}\matT{H}\inv{A}\mat{H}\inv{K_y}$.

%%%%%%%%%%%%%%%%%%%%%%%%%%%%%%%%%%%%%%%%%%%%%%%%
\section{Example curves} \label{sec:comparison}
%%%%%%%%%%%%%%%%%%%%%%%%%%%%%%%%%%%%%%%%%%%%%%%%

\begin{figure}[tbhp]
%\begin{center}
\hspace{-0.1in}\includegraphics[width=1.05\linewidth, keepaspectratio]{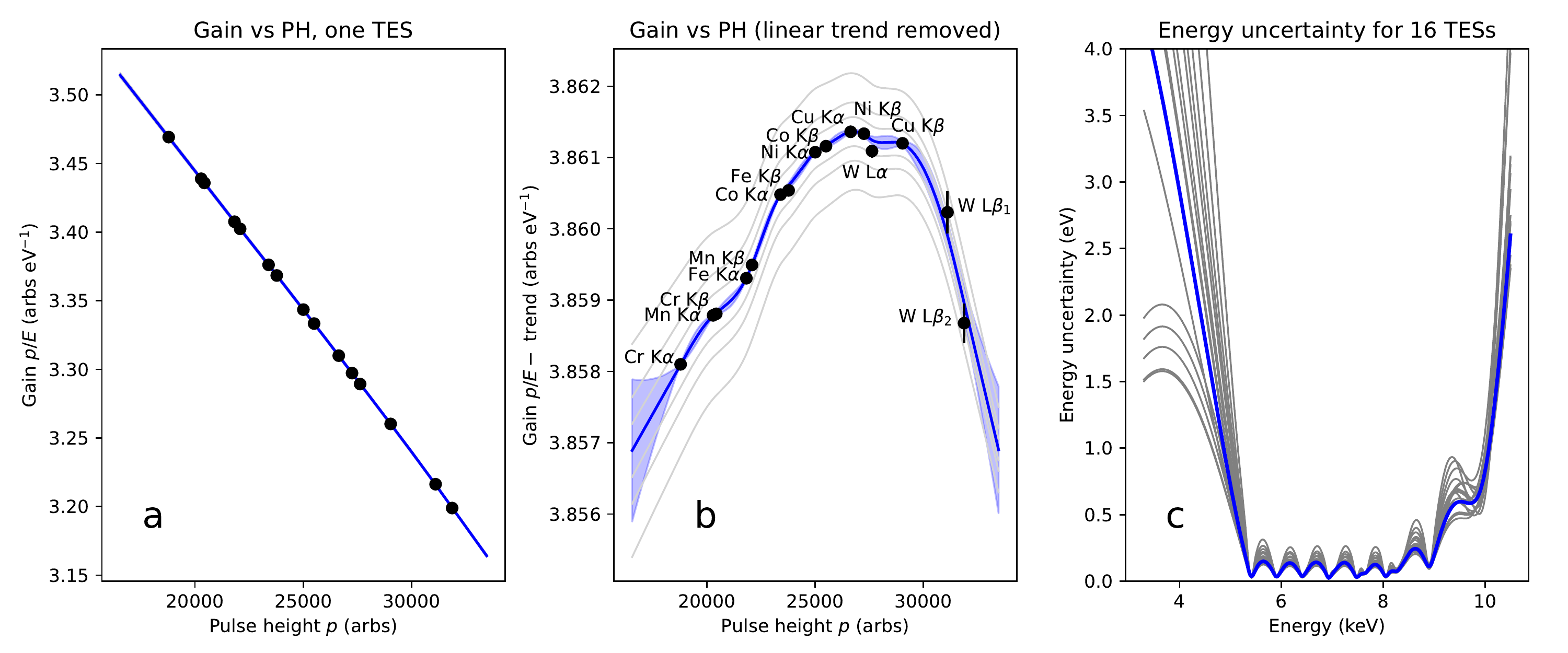}
\caption{\label{fig:figname}
{\it Panel a:} Example calibration curve (gain vs.\ PH) for one representative TES\@.
{\it Panel b:} Same data as panel a, except here the linear trend of $-2.07\times 10^{-5}\,\mathrm{eV}^{-1}$ is subtracted from gain, to highlight departures from the trend.  Labeled points ($\bullet$) with $1\sigma$ error bars are the 15 anchor points used. The solid curve is the smoothing spline (Equation~\ref{eq:margmean}), and the shaded band represents the $\pm1\sigma$ calibration uncertainty (square root of Equation~\ref{eq:variance}). Six thin gray curves are placed at $\pm 0.5$\,eV, $\pm 1$\,eV, and $\pm 2$\,eV about the best calibration to indicate the energy scale.
{\it Panel c:} The calibration curves' energy uncertainty as a function of energy for 16 representative TESs (the heavier line corresponds to the same TES featured in panels a and b). Anchor points range from 5.4\,keV to 10\,keV\@. The calibration uncertainty is lower closest to the anchor points, particularly those measured with low uncertainties. The five sensors with smaller uncertainties below 4\,keV are those that measured enough Si K$\alpha$ emission at 1.7\,keV to have an additional anchor point at that energy.
(Color figure online.)}
%\end{center}
\end{figure}

Figure~\ref{fig:figname} shows the results of one such calibration procedure, with a cubic smoothing spline of gain $g=p/E$ vs.\ $p$. The calibration curve is shown both directly and with a linear trend subtracted to emphasize the difference between the trend and the complete calibration curve. The figure also illustrates the varying GPR-estimated uncertainty associated with anchor points not uniformly spaced and with unequal uncertainties. Through most of the energy range populated by anchor points, 5.4\,keV to 9\,keV, the calibration uncertainty is well below 0.5\,eV, as are the small-scale features in the curve. As expected, the uncertainty is smallest at the well-measured anchor points and grows with distance from the nearest one. Changing the variance $\sigma_f^2$ by a factor of $2^{\pm 1}$ relative to its maximum-likelihood value moves energy estimates by no more than $\pm 0.1$\,eV in most cases.

%%%%%%%%%%%%%%%%%%%%%%%%%%%%%%%%%%%%%%%%%%%%%%%%
\section{Conclusion} \label{sec:conclusion}
%%%%%%%%%%%%%%%%%%%%%%%%%%%%%%%%%%%%%%%%%%%%%%%%

% 1. What we've done
We have considered the construction of energy-calibration curves for nonlinear TES microcalorimeters in the framework of Gaussian processes. With the once-integrated Wiener process as our specific GP model, the predictions follow a cubic smoothing spline with natural boundary conditions. We have previously shown that such an approximating spline is an excellent match to TES calibration~\cite{Fowler2021}, but the GPR framework also permits computation of the calibration uncertainties.

% 2. What we suggest
We can briefly summarize the calibration method:
\begin{enumerate}
    \item Using triples of calibration points, learn which calibration space has the least curvature for the current data (Section~\ref{sec:cal_space}).
    \item Find the maximum-likelihood GP variance, $\sigma_f^2$ (maximize Equation~\ref{eq:marglikely}).
    \item Compute the predictions $\overline{h}(\vec{x})$ for each anchor point (Equation~\ref{eq:margmean}).
    \item Calibration $h(x)$ is the cubic smoothing spline that interpolates those predictions.
\end{enumerate}
Item 1 should be done once for the given measurement and detector array; the other steps are repeated once for each calibration (say, once per sensor per day). We plan further investigation into whether it is better to use a separate $\sigma_f^2$ for each sensor or a single, universal value. We intend to use the calibration method based on GPR, and the energy uncertainty it generates, for future TES calibrations.

\begin{acknowledgements}
This work was supported by NIST's Innovations in Measurement Science program. We thank Dan Becker, Michael Frey, and two anonymous reviewers for many helpful suggestions.

The datasets generated during and/or analysed during the current study are available from the corresponding author on reasonable request.
\end{acknowledgements}

% Create the reference section using BibLateX:
\printbibliography

\end{document}